\documentclass[aip,reprint]{revtex4-1}

\usepackage{amsmath}
\usepackage{amssymb}
\usepackage{amsfonts}
\usepackage{color}
\usepackage{ulem}
\usepackage{graphicx}
\usepackage{natbib}
\begin{document}
\title{Generating photon pairs from a silicon microring resonator using an electronic step recovery diode for pump pulse generation}
\author{Marc Savanier}
	\email{msavanier@eng.ucsd.edu}
\author{Shayan Mookherjea}
	\email{smookherjea@eng.ucsd.edu}
\affiliation{Department of Electrical \& Computer Engineering, University of California, San Diego, La Jolla, CA 92093, USA}
\date{\today}

\begin{abstract}
Generation of photon pairs from compact, manufacturable and inexpensive silicon (Si) photonic devices at room temperature may help develop practical applications of quantum photonics. An important characteristic of photon-pair generation is the two-photon joint spectral intensity (JSI), which describes the frequency correlations of the photon pair. In particular, heralded single-photon generation requires uncorrelated photons, rather than the highly anti-correlated photons conventionally obtained under continuous-wave (CW) pumping. Recent attempts to achieve such a factorizable JSI have used short optical pulses from mode-locked lasers, which are much more expensive and bigger table-top or rack-sized instruments compared to the Si microchip pair generator, dominate the cost and inhibit the miniaturization of the source. Here, we generate photon pairs from a Si microring resonator by using an electronic step-recovery diode to drive an electro-optic modulator which carves the pump light from a CW optical diode into pulses of the appropriate width, thus potentially eliminating the need for optical mode-locked lasers.        
\end{abstract}

\maketitle

Room-temperature photon-pair and heralded single photon sources near $1.55\ \mathrm{\mu m}$ wavelength have traditionally been made using nonlinear crystals or optical fibers. Silicon (Si) photonic devices can also be used for photon-pair generation based on the spontaneous four-wave mixing (SFWM) nonlinearity.\citep{sharping2006generation, Harada2008,clemmen2009continuous} Unlike quantum dots, these photon sources can operate at room temperature and can be coupled relatively easily to optical fibers. The pump and generated photon wavelengths all lie within the telecommunications spectral windows used in fiber-optic networks. Moreover, such devices can nowadays be manufactured using CMOS-compatible processes on Si-on-insulator (SOI)\citep{davanco2012} and Si wafers,\citep{Gentry15} thus lowering the cost and increasing the likelihood for widespread applications.    

It has been shown that photon pairs can be created in simple-to-fabricate structures such as Si microring resonators at a rate of a few hundred kilohertz, using approximately a milliwatt of optical pump power, based on the process of SFWM. Although both Si waveguides and micro-resonators have been used to create photon pair-generation devices, resonators use the optical pump power more efficiently than waveguides and they have generated photon pairs with a higher coincidence-to-accidentals (CAR) ratio. \citep{harris2014integrated,grassani2014micrometer,Suo:15} The quality factors ($Q$) of Si microrings used for pair generation have recently improved,\citep{Savanier:16} but are generally lower than silicon nitride or glass microrings. However, silicon microrings have other benefits, including a compact footprint, a larger free-spectral range with fewer pairs of spectral lines being generated (since most applications today require only a single pair of photons), and since the pair-generation wavelengths shift with temperature,\citep{kumar2013spectrally} an electronic $p$-$i$-$n$ diode can be incorporated in the cross-section of the Si microring itself for real-time monitoring.\citep{savanier2015optimizing}

SFWM consists in the generation of two photons at radial frequencies $\omega_1$ and $\omega_2$ from two pump photons at $\omega_p$, with respective wavelengths $\lambda_i$ where $i = \{1, 2, p\}$, and energy conservation reading $2 \omega_p = \omega_1 + \omega_2$. In a microring resonator, its efficiency is maximum when the triple resonance condition is also fulfilled.

\begin{figure}
\includegraphics[width=\columnwidth]{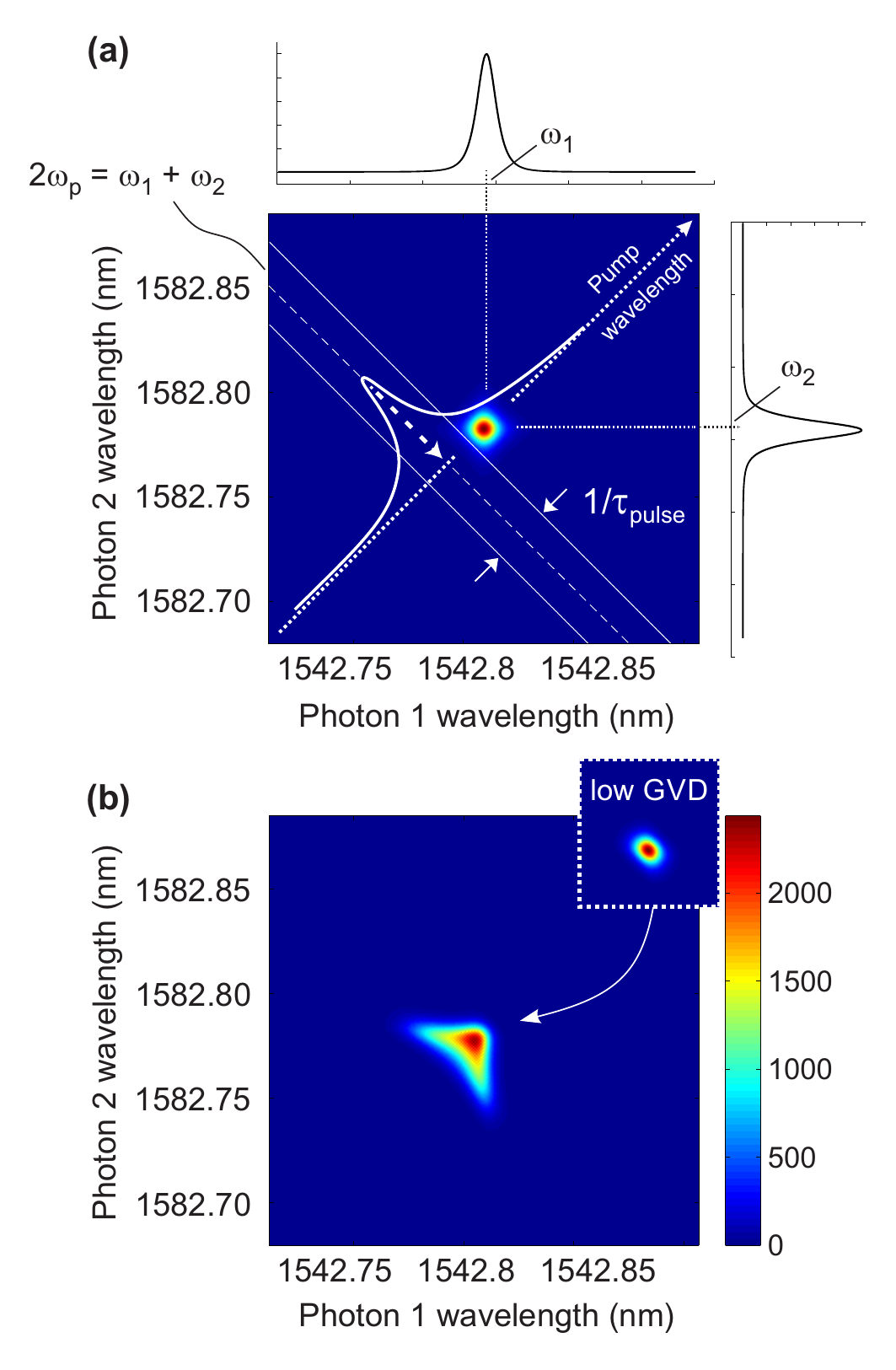}
\caption{(a) Calculated spectral properties of the photon pairs generated by the Si microring resonator used in these experiments show that the center of the joint spectral resonance of the photon pair (at radial frequencies $\omega_1$ and $\omega_2$) does not exactly coincide with the microring resonance at the pump frequency ($\omega_p = (\omega_1 + \omega_2)/2$), because of the fairly strong dispersion of Si nanophotonic waveguides (-1040~ps/nm-km at a wavelength of 1550 nm) which form the microring resonator (see Fig.~\ref{F2}). (b) Consequently, the JSI is expected to have the indicated shape; the inset shows the JSI that would be expected from a (hypothetical) waveguide cross-section with 10x lower group velocity dispersion.}
\label{F1}
\end{figure}

If a continuous-wave (CW) optical pump is used, the photon pairs that are generated by spontaneous nonlinear processes are highly correlated in the frequency-time domain, and can be entangled--as verified by two-photon interference measurements. The generated bi-photon spectrum, plotted in the ($\lambda_1, \lambda_2$) space and known as the joint spectral intensity (JSI) of the photon pair is generally elliptical (or multi-peaked), with a Schmidt number significantly greater than 1. When a heralded single-photon source is desired, it is necessary that the JSI be circular, which leads to a separable two-photon wavefunction.\citep{Mosley2008}  The width along the anti-diagonal axis in Fig.~\ref{F1}(a) (top-left to bottom-right) is determined by the product of the microring resonances at the generated photons wavelengths. The width of the JSI along the diagonal axis (from bottom-left to top-right) is determined by the convolution of the microring resonance at the pump wavelength and the pump pulse spectrum. Therefore, to achieve a circular JSI, we need to match the pump pulse duration to the resonance photon lifetime.

One approach is to use mode-locked lasers which generate short pulses optically;\citep{LPOR:LPOR201400404,2016arXiv160204962G}  however, these are large, expensive and individually-assembled laboratory instruments, and reduce the level of integration, scalability and cost-advantages of using integrated photonics. It would be preferable to obtain pulses from a CW laser diode using a standard amplitude electro-optic modulator driven by a compact, low-cost electrical pulse generator. 

\begin{figure*}
\includegraphics[width=\linewidth]{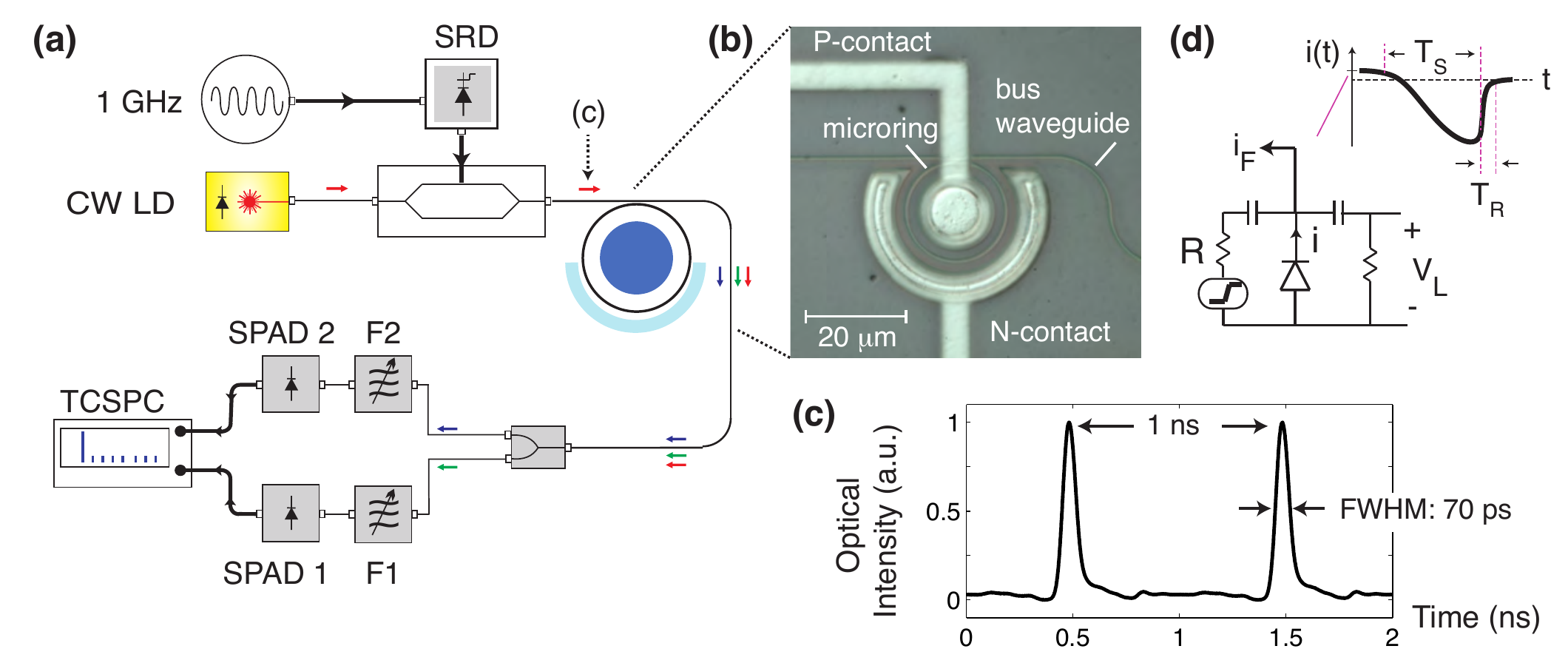}
\caption{(a) Experimental setup for pair generation and coincidence measurement using a waveguide-coupled microring resonator, with light from a continuous wave laser diode (CW LD) was carved into short pulses (width 63~ps) using a conventional lithium-niobate electro-optic modulator whose RF input was driven by a step-recovery diode (SRD). (b) A microscope image of the microring resonator with embedded diode for resonance monitoring. (c) The SRD was triggered by a sinusoidal waveform at 1 GHz, and drives a conventional electro-optic modulator to carve pulses of width 70~ps into the optical pump light from the CW LD, as measured by a wide-bandwidth oscilloscope (Agilent DCA-X with 30~GHz module) and optical detector (HP 83440D; impulse response $<$15~ps).}
\label{F2}
\end{figure*}

\textit{Device:}~Our device, shown in Fig.~\ref{F2} was fabricated at IME (Singapore) using CMOS-compatible process on standard SOI wafers widely used in Si photonics, with 220~nm Si thickness and $3\ \mathrm{\mu m}$ buried oxide thickness. Two etch steps were used to give a rib waveguide cross-section of height 220 nm, a waveguide width of 650 nm, and a slab thickness of 70 nm. The microring resonator had a radius of $R$ = 10 $\mu$m with inter-waveguide gap of 200 nm (satisfying the minimum feature-size recommendation for the foundry fabrication process). Only the lowest-order transverse electric (TE) polarized mode family propagates in such microrings.  The $p$ and $n$ doped regions were formed by implanting Boron and Phosphorous, respectively, with the edge of the implanted regions being 900~nm from the edge of the ridge. These implanted regions create a photodiode which monitors the optical power in the microring, so that the pump wavelength can be aligned to the microring resonance despite temperature variations, as discussed elsewhere.\citep{savanier2015optimizing} Test waveguides similar to those used to form the microring resonator were measured using an atomic force microscope to have a root-mean-squared sidewall roughness of 2.6~nm, and, using an optical cutback method, to have a propagation loss of -0.74 dB/cm at 1550~nm. Using the rib waveguide cross-section, we can achieve a more favorable group velocity dispersion (GVD) than the fully-etched Si waveguide. We calculated a GVD coefficient of -1040 ps/nm-km at a wavelength of 1550 nm, whereas the ``standard'' fully-etched Si waveguide, with a similar cross-section has a much higher group-velocity dispersion (4400 ps/nm-km),\citep{Dulkeith:06} as well as a higher optical propagation loss (typically 2--4 dB/cm)\citep{Vlasov:04,6324087}, both of which can negatively impact pair generation. Fig.~\ref{F1}(b) suggests that we should continue to attempt to further lower the GVD from what we report here.  

\textit{Pulse generation:}~Here, we study photon pairs generated by a silicon microring device, pumped optically by CW light from a laser diode which is carved into pulses using an off-the-shelf electro-optic amplitude modulator driven by a compact and inexpensive step-recovery diode (SRD).\citep{appnote-SRD} An SRD is a two-terminal $p$-$i$-$n$ diode whose junction impedance depends upon its internal charge storage. Fig.~\ref{F2}(d) sketches the operating principle of the SRD. Under forward-bias, the junction has low impedance and charge build-up occurs. When driven by the portion of the sinusoidal RF drive which reverse-biases the diode, the junction remains low-impedance until the stored charge is depleted. This condition abruptly causes the impedance to rise, stopping the flow of current. The forward bias current $i_F$, the equivalent resistance seen by the diode $R_\text{eq} = R/2$, the desired peak voltage at the output $V_L$ and the minority carrier lifetime $\tau$ determine the charge storage time $T_S \approx (i_F R_\text{eq} /V_L) \tau \approx 0.5~\text{ns}$ in our experiment. The voltage pulse is generated during the (much shorter) rise (``recovery'') time of the diode current ($T_R$), which is determined by $R_\text{eq}$ and the diode junction capacitance $C_J$, so that $T_R \approx 3.1 R_\text{eq} C_J$, i.e.,~generating a pulse of duration 70~ps requires $C_J \approx 1$~pF. Although we use a commercial SRD module in this demonstration (Herotek GIM 1000A; bias current approximately 200~mA at 5~V DC bias, generating 9~V peak-voltage pulses of duration 63~ps into a 50~$\Omega$ load), such diodes can be fabricated on a chip, and can generate pulses of a few tens of picoseconds, if care is taken to minimize parasitic inductances and capacitances which limit the voltage slew-rate.     

It is important to note that the specifications of the SRD guide the design of the Si microring resonator, and vice-versa. The relatively low-Q (unloaded quality factor $Q_U<1 \times 10^5$) Si microrings used for pair generation until recently\citep{Savanier:16} require pulses that are only a few picoseconds in duration, and simple all-electronic pulse generation approaches are impossible. On the other hand, very high-Q Si micro-resonators can be fabricated,\citep{borselli2005beyond} but create challenges in stabilization of the pair generation rate (PGR) and fabrication yield, because a phase-matching `sinc' function comes into play, and small fabrication imperfections or environmental variations can cause the PGR to oscillate between 0 and its design value.\citep{Savanier:16} We operate in the intermediate sweet-spot ($Q_U \approx 1 \times 10^6$), where the microring-waveguide coupling coefficient is chosen such that the loaded-cavity ($Q_L \approx 1 \times 10^5$) ringdown time is more or less matched to the SRD pulsewidth.    

\textit{Experiment:}~The microresonator was optically pumped in the usual experimental setup for SFWM.\citep{KumarOpEx15} An optical CW diode laser was carved into optical pulses at a repetition rate of 1~GHz using an off-the-shelf, fiber-pigtailed high bandwidth (40~GHz) amplitude electro-optic (EO) lithium-niobate modulator. To achieve a high extinction ratio at the modulator [see Fig.~\ref{F2}(c)], the EO modulator bias voltage was continuously controlled using a micro-electronic circuit based on a small pilot tone at the null-mode operating point. The RF input to the modulator was generated by the SRD. Based on the measured optical linewidth of the microring resonance, the photon lifetime was estimated to be $79\ \mathrm{ps}$. An SRD with a slightly shorter pulsewidth was used to drive the EO modulator [see Fig.~\ref{F2}(c)], allowing for a slight pulse broadening through the fibers and feeder waveguide.   

\begin{figure*}
\includegraphics[width=\linewidth]{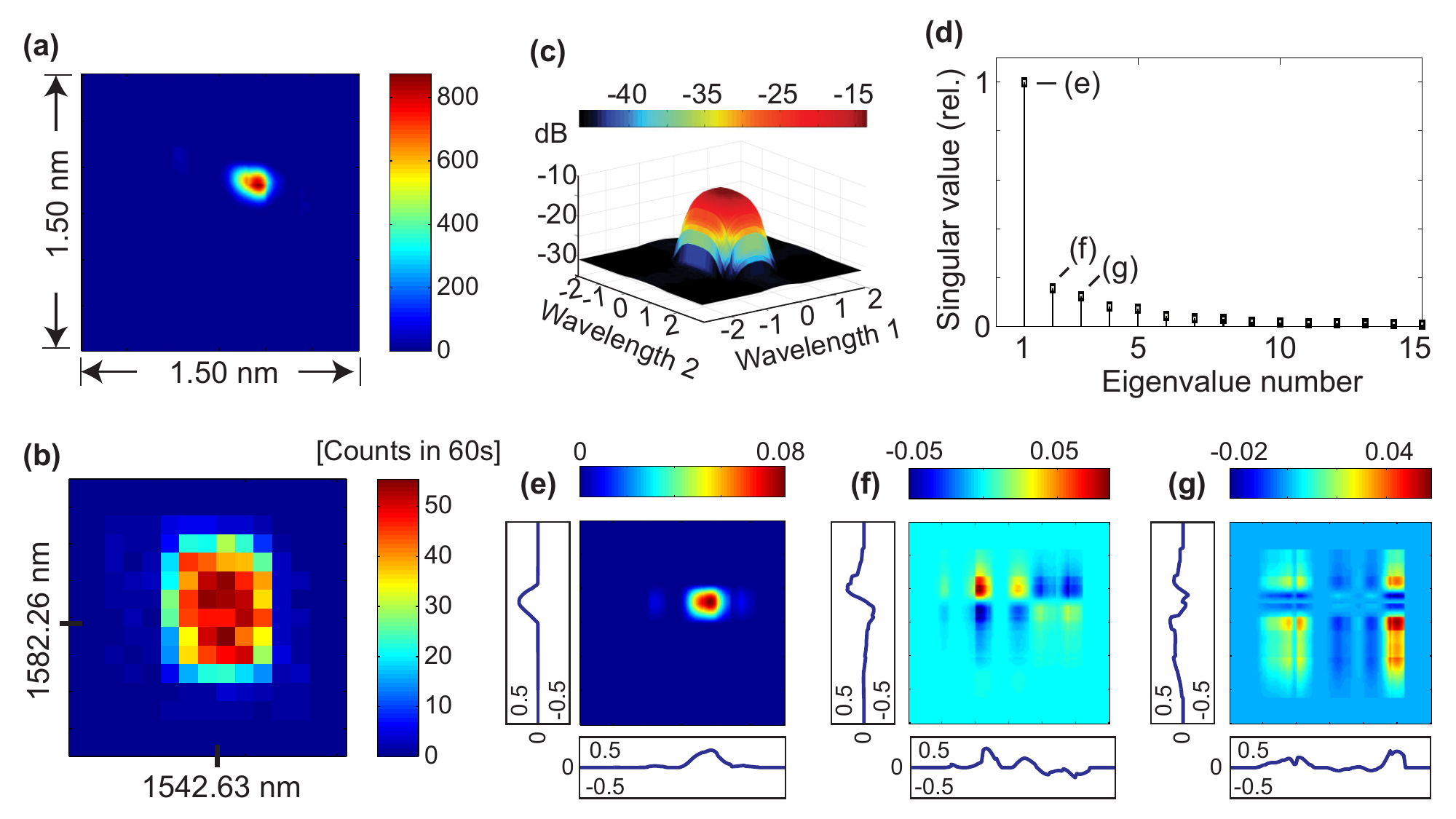}
\caption{(a) De-convolved experimental JSI, and (b) raw measured JSI, which is convolved with (c) the point spread function (PSF) of the filter assembly. Tunable optical filters were used in front of the single-photon avalanche detectors (SPADs) to measure the JSI. The filter point-spread function (PSF) had a full-width at half-maximum of 0.6 nm along the axis for photon 1 and 1.0 nm along the axis for photon 2. (d) Distribution of the singular values of the JSI, dominated by the contribution of the first mode and a few others. (e)-(f) The first three modes in the singular value decomposition of the JSI.}
\label{F3}
\end{figure*}

Based on a separate calibration measurement, the fiber-to-waveguide optical insertion loss was estimated to be -3.5~dB at each facet. Photons at the output of the chip were spectrally separated using a custom-built tunable filter assembly (using telecommunications components) that suppressed the residual pump photons by more than 150~dB and extracted the signal and idler wavelengths, with insertion loss -6~dB and -5~dB, respectively. Single photons at the signal and idler wavelengths were detected using gated thermo-electrically cooled (233~K) InGaAs SPADs with an estimated 10$\%$ quantum efficiency, electrically-generated gate width of 3.0~ns and gating repetition frequency 60~MHz \citep{Tosi-2012}. The dark counts of the two SPADs were 1.4~kHz and 3.0~kHz. A time-correlated single photon counting (TCSPC) board was used to measure histograms of coincidence and accidental counts. The SPAD controller and the RF oscillator used to drive the SRD were sychronized at 10~MHz. The gate window at the SPADs accommodates 3 pump pulse time bins, each of which has a low generation probability $\approx$ 5.1 $\times$ 10$^{-5}$ pairs/pulse.  Under these operating conditions, at a temperature of 30$^\circ$C and average input pump power (in the feeder waveguide) of 1~dBm at 1562.209~nm, the photon pairs at 1542.63~nm and 1582.26~nm were generated at a rate of about 733 kHz. 

\textit{JSI measurement:}~Measurement of the JSI shown in Fig.~\ref{F3} is time-consuming, and required the acquisition of 225 histograms of the start-stop coincidence inter-arrival times, each integrated for 60 seconds. (A classical seeded four-wave mixing experiment more rapidly measures the phase-matching properties inherent to the spontaneous mixing process\citep{eck13,Fang:14}, although not directly measuring the device when it is generating photon pairs.) To acquire the JSI more rapidly, we operated with a higher average pump power than we usually do; at lower pump powers, a $\text{CAR} \sim 100$ has been measured from the same microrings.\citep{savanier2015optimizing} The optical intensity enhancement factor (ratio of the circulating field intensity in the microring to the intensity in the feeder waveguide) was 346. Thus, an input waveguide power of 1~dBm resulted in a circulating field intensity in the microring of about $240\ \mathrm{MW}.\mathrm{cm}^{-2}$. Under these conditions, the associated two-photon absorption in silicon was calculated to be 0.5~dB/cm, i.e.~slightly smaller than the linear propagation loss. In addition, from the higher pump power, a wavelength red-shift of 48~pm was observed, relative to the lower average pump power results we reported earlier.\citep{savanier2015optimizing}

 The measured ``raw'' JSI represents the convolution of the actual JSI and the separately-measured (static) point-spread function of the filters. Similar to our previous work,\citep{art-Kumar-NCOMM-2014} we used the damped Richardson-Lucy (RL) algorithm,\citep{bigg97} an iterative procedure for recovering, in a maximum-likelihood sense, a latent image that has been blurred by a known point spread function (PSF). The end-point of the RL iteration generally needs to be determined by the user, and we confirm our choice by a classical (seeded) four-wave mixing experiment on the same device. Using 25 iterations and a damping value of 1 standard-deviation (which are fairly typical values), the result is shown in Fig.~\ref{F3}(a). The `raw' blurred JSI is shown in Fig.~\ref{F3}(b), and the PSF of the optical filters in Fig.~\ref{F3}(c). The PSF of these filters can be measured very accurately using a tunable CW laser and high dynamic range photodetector. The reconstruction does not show significant sensitivity to the exact number of iterations, or the value of the damping parameter. We used a sub-sampling procedure (by 8x) during reconstruction, which yields better positional fidelity when the PSF is well known, and results in a more accurate determination of lateral shifts of the reconstructed image in the plane.\citep{chap-Hanisch} This allowed us to confirm the consequence of the wavelength red-shift discussed above in causing the experimentally-measured JSI [Fig.~\ref{F3}(a)] to move to longer wavelengths (towards the upper-right corner) compared to the simulated JSI [Fig.~\ref{F1}(b)], which ignored this thermal effect in the calculation. 

\textit{Schmidt decomposition:}~A quantitative measure of the factorizability of the generated two-photon states is obtained by performing a Schmidt decomposition.\citep{law00} A bi-partite state is separable (and suitable for heralding) if the number of non-zero eigenvalues is one. A widely-used operational measure for entanglement is therefore obtained by calculating the Schmidt number $K$, representing the number of orthogonal modes in the singular-value decomposition (SVD) of the magnitude of the joint spectral amplitude (JSA, calculated here as the square root of the measured JSI under the assumption of flat spectral phase). For the measured JSI shown in Fig.~\ref{F3}(a), we calculate $K = 1.12$, whereas the theoretically-expected value from Fig.~\ref{F1}(d) is $K = 1.21$. This is already a good level of agreement; furthermore, the simulation relies on the assumption of a specific waveguide height and width, whereas the fabricated waveguide inevitably had slightly different dimensions, which could alter the JSI as shown in the inset to Fig.~\ref{F1}(b). 

Fig.~\ref{F3}(d) shows the distribution of the magnitudes of the first few singular values of the measured (deconvolved) JSI. This shows that the first mode dominates the distribution, although the contribution of the other modes is not zero. The shapes of the first three mutually-orthogonal spectral basis functions that form the singular value decomposition (SVD) of the JSA, are shown in Fig.~\ref{F3}(e)-(g), and as expected, show the symmetrical or anti-symmetrical behavior about the spectral center-point. Elimination of the contribution of the higher-order modes can be acheived by reducing the degree of spectral asymmetry shown in Fig.~\ref{F1}(a) between the pump (thick dashed white arrow line) and the center spot of the dual photon resonance.   

\textit{Summary:}~Previously, generating close-to-circular JSI's with a low Schmidt number required precision poling of nonlinear crystals or pump beam shaping by bulk optics components, and the use of mode-locked lasers which generate short optical pulses. Here, the SRD solves the problem of pump-pulse generation in a compact footprint, and requires only a sinusoidal electronic drive waveform. The remaining impediment towards a completely circular JSI is the waveguide dispersion, which is quite large in conventional Si photonic structures, but can be improved with a new round of fabrication. More generally, our work demonstrates that Si microfabrication can be used to create simple microring based room-temperature optically-pumped pair generators. In the future, we anticipate that the Si micro-electronic components and the electro-optic modulator, and perhaps also the laser diode itself, can all be co-integrated with the Si microring resonator onto a monolithic platform, for a truly integrated photon-pair source. 

\vspace{6pt}
The authors are grateful to R. Kumar (Intel Research, Santa Clara, CA), J. R. Ong (Institute of High Performance Computing, ASTAR, Singapore),  Xianshu Luo and Guo-Qiang Patrick Lo (Institute of Microelectronics, ASTAR, Singapore), Nick Bertone (Optoelectronic Components), and the National Science Foundation for support (ECCS 1201308). 


%

\end{document}